\documentclass[journal]{IEEEtran}
\usepackage{amsmath,amssymb,amsfonts}
\usepackage{graphicx}
\usepackage{cite}
\usepackage{bm}
\usepackage{mathtools}
\usepackage{booktabs}
\usepackage{needspace}
\usepackage{microtype}
\newcommand{\indicator}{\mathbf{1}}
\DeclareMathOperator{\osc}{osc}
\DeclareMathOperator{\clip}{clip}
\newtheorem{theorem}{Theorem}
\newtheorem{proposition}{Proposition}

\begin{document}
\raggedbottom
\setlength{\textfloatsep}{12pt plus 2pt minus 2pt}
\setlength{\dbltextfloatsep}{12pt plus 2pt minus 2pt}
\setlength{\jot}{2pt}

\title{Queue-Aware ISAC Power Allocation Under Imperfect CSI: Integer-Service Analysis and Anchored Lyapunov Look-Ahead}

\author{Hanyi~Sun, Yifan~Zhou, Yixiao~Gu,~\IEEEmembership{Member,~IEEE}%
\thanks{The authors are with the School of Communication and Information Engineering, Shanghai University, Shanghai 200444, China (e-mail: sunhanyi05@shu.edu.cn; zhouyifan@shu.edu.cn; yxgu@shu.edu.cn). \textit{(Corresponding author: Yixiao Gu.)}}}

\maketitle

\begin{abstract}
We study power allocation for a multiuser integrated sensing and communication (ISAC) downlink with bursty queues and imperfect channel state information (CSI). Starting from a long-run utility--backlog objective, we derive its common task reward and online score. Integer departures are represented by threshold events of indefinite Gaussian quadratic forms, accounting for estimation error in both signal and interference. Their moments give the conditional one-slot task reward and characterize the error of an efficient plug-in score. An anchored quadratic Lyapunov look-ahead corrects its reward-greedy action with a bounded nominal loss. Across five CSI-error levels, the rule reduces mean queue exceedance by 27.4--43.9\% (0.49--0.99 percentage points) relative to the plug-in maximizer, with mean reward loss below 0.01 per slot. The rule achieves higher reward and lower exceedance than DPP, SAC and PPO; its advantage over freshly retrained learning baselines extends to five configurations with 9--23 power-allocation variables using fixed candidate budgets. For the seven-variable system, a fused implementation reproduces all 25 confirmation trajectories and achieves a one-core p99 decision time of 0.274--0.311\,ms.
\end{abstract}


\section{Introduction}
Integrated sensing and communication (ISAC) shares spectrum, power and hardware~\cite{ref1_liu}. With bursty traffic and imperfect channel estimates, powers chosen from estimated channels produce uncertain integer packet departures. A favorable aggregate reward can also leave individual queues congested. These two effects motivate a reward-aligned rule with explicit control of its queue correction.

Beamforming optimizes communication and sensing metrics~\cite{ref5_hua,ref7_liu_cramer}, while Lyapunov DPP~\cite{ref13_neely} addresses long-term constraints. Zakeri \textit{et al.}~\cite{ref_zakeri} consider single-user, single-target precoding with an average-SINR virtual queue. Yang \textit{et al.}~\cite{ref_yang_dynamic} allocate power among vehicular V2I, V2V and sensing beams, using traffic queues and sensing mutual information. We consider a multiuser BS downlink where bursty communication and sensing-task queues share power under imperfect CSI. Here, the same channel error perturbs signal and all interference, changing integer departures and the next-slot backlog penalty.

Expected-service scheduling~\cite{ref_ouyang}, OFDM delay--power--CSIT analysis~\cite{ref_lau} and imperfect-CSI rate adaptation~\cite{ref_hu} establish the conditional-service approach. DRL supports ISAC allocation~\cite{ref10_yang,ref11_ye,ref15_qin}; Lyapunov-guided learning uses DPP decisions as targets~\cite{ref14_bi} or its negative objective as the SAC reward~\cite{ref17_liang}. Our objective combines communication--sensing utility with aggregate backlog cost. Integer-service moments give its conditional reward; an efficient plug-in score then provides the anchor for a queue-distribution correction. The contributions are:
\par\noindent 1) Full-SINR integer-service analysis maps CSI uncertainty to departure moments and conditional task reward, with Cantelli bounds and a convergent spectral probability series.
\par\noindent 2) An anchored quadratic backlog correction refines the reward-aligned scheduler. Its nominal loss is at most $\rho$; the conditional loss adds the score-error oscillation, following the $h$-myopic principle~\cite{ref_meyn}.
\par\noindent 3) Common-stream experiments cover CSI errors, traffic variations and 9--23 allocation variables, with freshly retrained learning baselines and a trajectory-verified CPU implementation.

\section{System Model and Control Objective}
\label{sec:system}
An $N$-antenna BS serves $M$ single-antenna users and $K$ sensing targets in slots of length $\Delta_t$. It transmits $\mathbf x=\mathbf W_c\mathbf s_c+\mathbf W_s\mathbf s_s$ with total power at most $P_{\max}$.

\subsection{Communication Service}
\label{sub:comm}
User $m$ has channel $\mathbf h_m=\hat{\mathbf h}_m+\mathbf e_m$, where $\hat{\mathbf h}_m$ is the MMSE estimate and $\mathbf e_m\sim\mathcal{CN}(\mathbf0,\sigma_e^2\mathbf I_N)$ is independent of it and of other users' errors. Here $\sigma_e=\sqrt{\sigma_e^2}$ and $\sigma_e^2=1/(1+T_p\bar\gamma_p)$ for $T_p$ pilots at SNR $\bar\gamma_p$. The realized SINR is
\begin{equation}
\gamma_m=\frac{|\mathbf h_m^H\mathbf w_{c,m}|^2}{\sum_{j\ne m}|\mathbf h_m^H\mathbf w_{c,j}|^2+\sum_k|\mathbf h_m^H\mathbf w_{s,k}|^2+\sigma_c^2}.
\label{eq:sinr}
\end{equation}
Replacing $\mathbf h_m$ by $\hat{\mathbf h}_m$ gives $\hat\gamma_m$. With $\zeta=B\Delta_t/(L_p\ln2)$, offered service is $\mu_m^c=\lfloor\zeta\ln(1+\gamma_m)\rfloor$ packets of $L_p$ bits; the BS predicts $J_m=\lfloor\zeta\ln(1+\hat\gamma_m)\rfloor$. This is capacity-type service under realized SINR, with estimated-CSI decisions; decoding errors and retransmissions are not modeled.

\subsection{Sensing Service and Traffic Queues}
\label{sub:sensing}
Target $k$ has response $\mathbf G_k=\varsigma_k\alpha_k\mathbf a_R(\theta_k)\mathbf a_T^H(\theta_k)$, with independent per-slot $\alpha_k\sim\mathcal{CN}(0,1)$. After communication-clutter suppression, the BS knows $\eta_k=\|\mathbf G_k\mathbf W_s\|_F^2/\sigma_s^2$, and serves $\mu_k^s=\mu_0\indicator\{\eta_k\ge\eta_{\min}\}$ tasks.

All $M+K$ queues\label{para:queues} have unbounded buffers and update as $Q_i(t+1)=Q_i(t)-D_i(t)+A_i(t)$, where $D_i=\min\{Q_i,\mu_i\}$. Each user has an independent hidden two-state Markov mode setting its Poisson arrival rate (MMPP); sensing arrivals are independent Poisson$(\lambda_s)$. Arrivals are independent of current channel errors. Exceedance is the fraction of slots in which any queue exceeds $Q_{\rm th}$.

\subsection{Objective, Reward and Power Allocation}
\label{sub:objective}
Let $U_t=\alpha\widetilde R_t+\beta S_t$ combine communication and sensing utility, and let $C(q)=\nu_1q+\nu_2q^2$ price backlog. We seek a causal policy $\pi$ maximizing
\begin{equation}
\max_{\pi\in\Pi}\ \liminf_{T\to\infty}\frac1T\sum_{t=0}^{T-1}
\mathbb E_\pi[VU_t-\delta Z_t-C(\bar q_{t+1})],
\label{eq:objective}
\end{equation}
where $\Pi$ uses pre-service queues, estimated channels, sensing responses and past arrivals, denoted $\mathcal F_t$. The per-slot contribution defines the common task reward
\begin{equation}
r_t=VU_t-\delta Z_t-\nu_1\bar q_{t+1}-\nu_2\bar q_{t+1}^{\,2}.
\label{eq:reward}
\end{equation}
Here $\bar q_{t+1}=\sum_i\varpi_iQ_i(t+1)$, with $\varpi_i=(2MQ_{\rm th})^{-1}$ for users and $(2KQ_{\rm th})^{-1}$ for targets. The normalized rate is $\widetilde R=\sum_m\log_2(1+\omega_m\hat\gamma_m)/(M\log_2(1+N)+10^{-6})$, where $\omega_m=\|\hat{\mathbf h}_m\|^2/(\|\hat{\mathbf h}_m\|^2+N\sigma_e^2+10^{-12})$. The sensing success fraction is $S$; $Z$ is the mean normalized shortfall $(1-\eta_k/\eta_{\min})_+$.

MRT directions for users and unit-norm steering directions for targets are fixed. An action $\mathbf a\in[-1,1]^{M+K}$ sets $p_i=P_{\max}e^{a_i}/\sum_\ell e^{a_\ell}$. At fixed power fractions, every SINR and sensing SNR is nondecreasing in total power; service and reward increase or stay unchanged. All methods therefore use full power. Online control uses the conditional one-slot reward $G(\mathbf a)=\mathbb E[r_t\mid\mathcal F_t,\mathbf a]$ derived below to construct an efficient score and a bounded queue correction.

\section{Conditional Service and Reward}
\label{sec:theory}
\subsection{Integer-Service Thresholds}
\label{sub:thresholds}
For a fixed action, set $\mathbf B_m=\sum_{u\ne m}\mathbf w_{c,u}\mathbf w_{c,u}^H+\sum_k\mathbf w_{s,k}\mathbf w_{s,k}^H$, $n=\sigma_c^2>0$, $\tau_j=e^{j/\zeta}-1$ and $\mathbf H_{mj}=\mathbf w_{c,m}\mathbf w_{c,m}^H-\tau_j\mathbf B_m$. Write $p_{ij}=\Pr(\mu_i\ge j\mid\mathcal F_t,\mathbf a)$ and $\hat J_i=\min\{Q_i,J_i\}$, with $J_i=\mu_i^s$ for targets. The current errors are independent of $\mathcal F_t$. All moments below condition on this information and the action.

\begin{theorem}[Integer departures under imperfect CSI]
\label{thm:departures}
Communication service obeys the full-SINR threshold equivalence
\begin{equation}
\{\mu_m^c\ge j\}=\{X_{mj}\ge0\},\quad
X_{mj}=\mathbf h_m^H\mathbf H_{mj}\mathbf h_m-\tau_jn.
\label{eq:threshold}
\end{equation}
Its mean and variance are
\begin{equation}
\begin{aligned}
\bar x_{mj}&=\hat{\mathbf h}_m^H\mathbf H_{mj}\hat{\mathbf h}_m+\sigma_e^2\operatorname{tr}\mathbf H_{mj}-\tau_jn,\\
s_{mj}^2&=2\sigma_e^2\|\mathbf H_{mj}\hat{\mathbf h}_m\|^2+\sigma_e^4\operatorname{tr}(\mathbf H_{mj}^2).
\end{aligned}
\label{eq:threshold_moments}
\end{equation}
For every queue, the actual-departure moments are
\begin{equation}
\mathbb E[D_i]=\sum_jp_{ij},\qquad
\mathbb E[D_i^2]=\sum_j(2j-1)p_{ij},
\label{eq:moments}
\end{equation}
where both sums run over $j=1,\ldots,Q_i$. For communication, $p_{mj}$ has a spectral series with a computable truncation remainder; for sensing it is deterministic. At $\sigma_e^2=0$, $p_{mj}=\indicator\{j\le J_m\}$.
\end{theorem}
Rearranging the full SINR proves~\eqref{eq:threshold}; integer tail sums give~\eqref{eq:moments}. Appendix~\ref{app:service} derives~\eqref{eq:threshold_moments}, the Cantelli error bound and the spectral series~\cite{ref_naffouri}.

\subsection{Conditional Reward and Online Score}
\label{sub:conditional}
Define $\lambda_i=\mathbb E[A_i\mid\mathcal F_t]$, $\varphi=\sum_i\varpi_i(Q_i+\lambda_i-\mathbb E[D_i])$, $\sigma_A^2=\operatorname{Var}(\sum_i\varpi_iA_i\mid\mathcal F_t)$ and $v_D=\sum_i\varpi_i^2\operatorname{Var}D_i$.
\begin{proposition}[Exact conditional reward]
\label{prop:interface}
The reward in~\eqref{eq:reward} satisfies
\begin{equation}
G(\mathbf a)=VU-\delta Z-\nu_1\varphi-\nu_2(\varphi^2+\sigma_A^2+v_D).
\label{eq:interface}
\end{equation}
\end{proposition}
Conditionally independent departures give $\mathbb E\bar q=\varphi$ and $\operatorname{Var}\bar q=\sigma_A^2+v_D$, establishing~\eqref{eq:interface}. The MMPP arrival mean $\lambda_i$ follows a causal forward filter. To avoid evaluating all service probabilities online, use stationary means $\bar\lambda_i$ and predicted departures in
\begin{equation}
\begin{aligned}
J_0(\mathbf a)&=VU-\delta Z-\nu_1\hat\varphi-\nu_2\hat\varphi^2,\\
\hat\varphi&=\sum_i\varpi_i(Q_i+\bar\lambda_i-\hat J_i).
\end{aligned}
\label{eq:plugin}
\end{equation}
Thus $J_0$ approximates the objective's one-slot contribution. For $e=G-J_0$ and $d_\varphi=\varphi-\hat\varphi$,
\begin{equation}
e=-[\nu_1+\nu_2(\varphi+\hat\varphi)]d_\varphi-\nu_2(\sigma_A^2+v_D).
\label{eq:error}
\end{equation}
Appendix~\ref{app:score} bounds these service errors using Theorem~\ref{thm:departures}. The same probabilities also give next-slot exceedance:
\begin{equation}
\begin{aligned}
\mathcal P_i(\mathbf a)={}&\bar F_i(Q_{\rm th}-Q_i)\\
&-\sum_{j=1}^{Q_i}b_i(Q_{\rm th}-Q_i+j)p_{ij}(\mathbf a),
\end{aligned}
\label{eq:risk}
\end{equation}
where $b_i(k)=\Pr(A_i=k\mid\mathcal F_t)$ and $\bar F_i(k)=\Pr(A_i>k\mid\mathcal F_t)$. Independence gives any-queue risk $\mathcal P=1-\prod_i(1-\mathcal P_i)$; its derivation is in Appendix~\ref{app:score}.

\section{Anchored Lyapunov Look-Ahead}
\label{sec:rule}
\subsection{Reward Anchor and Queue Correction}
\label{sub:correction}
The aggregate penalty in~\eqref{eq:reward} can assign equal costs to different queue distributions. To distinguish concentrated backlog, define post-decision queues~\cite{ref_powell} $y_i(\mathbf a)=Q_i-\hat J_i(\mathbf a)$ and potential $h(\mathbf y)=\tfrac12\sum_i(y_i/Q_{\rm th})^2$. Following $h$-myopic control~\cite{ref_meyn}, the potential guides a correction to the reward-greedy anchor $\mathbf a_0\in\arg\max_{\mathbf a\in\mathcal A}J_0(\mathbf a)$:
\begin{equation}
\begin{aligned}
d_\rho(\mathbf a)&=\clip\big(\chi[h(\mathbf y(\mathbf a_0))-h(\mathbf y(\mathbf a))],-\rho,\rho\big),\\
\mathbf a_1&\in\arg\max_{\mathbf a\in\mathcal A}\{J_0(\mathbf a)+d_\rho(\mathbf a)\}.
\end{aligned}
\label{eq:rule}
\end{equation}
Here $\clip(x,l,u)=\min\{u,\max\{l,x\}\}$; both maximizations select the first maximizer. The parameters $\chi\ge0$ and $\rho\ge0$ control the correction weight and nominal loss budget. The reward objective supplies $J_0$, while $h$ redistributes service toward larger residual queues.

\subsection{Loss Control and Candidate Selection}
\label{sub:loss}
\begin{proposition}[Anchored one-slot loss]
\label{prop:anchor}
On the same finite candidate set, let $\delta_0=J_0(\mathbf a_0)-J_0(\mathbf a_1)$ and $\osc(e)=\max e-\min e$. Then
\begin{equation}
0\le\delta_0\le\rho,\qquad
\max_{\mathbf a}G(\mathbf a)-G(\mathbf a_1)\le\delta_0+\osc(e).
\label{eq:transfer}
\end{equation}
\end{proposition}
Since $d_\rho(\mathbf a_0)=0$, maximization gives $J_0(\mathbf a_1)+d_\rho(\mathbf a_1)\ge J_0(\mathbf a_0)$ and hence $\delta_0\le\rho$. Substituting $G=J_0+e$ proves the conditional bound. At $\rho=0$ the rule is $\pi_0$; $\rho=\infty$ gives $\arg\max(J_0-\chi h)$.

The base system uses $\mathcal A=\{-1,0,1\}^7$; larger systems use a prescribed pool $\mathcal C$. Relative to the complete grid, nominal loss is at most $\varepsilon_{\mathcal C}+\rho$, where $\varepsilon_{\mathcal C}=\max_{\mathcal A}J_0-\max_{\mathcal C}J_0$. Before validation, we fixed $\rho=0.5$ and $\chi\in\{0.99,9.9,99,990\}$. Validation selects a common $\chi$ minimizing exceedance subject to mean reward loss $\le0.01$ at every CSI level and the growth check; ties use reward then smaller $\chi$. It selects 99: peak mean losses are 0.00924 for 99 and 0.01909 for 990.

\section{Performance Evaluation}
\label{sec:eval}
\subsection{Setup and Comparison Protocol}
\label{sub:setup}
The base setting is $N=8$, $(M,K)=(4,3)$, $B=100$\,MHz, $P_{\max}=1$\,W, $L_p=1000$ bits, $\Delta_t=0.5$\,ms, $\sigma_c^2=\sigma_s^2=10^{-10}$\,W, $\eta_{\min}=10$\,dB and $\mu_0=5$. Reward parameters are $(\alpha,\beta,\delta)=(0.5,1,10)$, $V=10$, $(\nu_1,\nu_2)=(5,20)$ and $Q_{\rm th}=100$. Rayleigh channels have $\mathbf h_m\sim\mathcal{CN}(\mathbf0,\mathbf I_N)$ and $\hat{\mathbf h}_m\sim\mathcal{CN}(\mathbf0,(1-\sigma_e^2)\mathbf I_N)$; the normalized downlink is predominantly interference-limited. Targets at $(-30,0,25)^\circ$ have $\varsigma_k^2=(1,1.2,0.9)10^{-7}$. Channels and target gains renew each slot. MMPP rates are 30/65 packets/user/slot, with normal-to-burst/reverse probabilities 0.05/0.1; mean arrivals are 125/3 per user and $\lambda_s=3$ per target.

Each CSI level uses three validation and five 10000-slot confirmation streams from empty queues. All methods share channels, errors, target gains and arrivals. DPP maximizes $\sum_iQ_iJ_i+\kappa_d(VU-\delta Z)$, selecting $\kappa_d=10^4$ among $10^0,\ldots,10^5$ by validation reward. Our original RLD-SAC design uses SAC~\cite{ref9_sac}; clipped-surrogate PPO follows~\cite{ref_ppo}. Both use three 256-unit Gaussian--tanh actor layers, estimated channels, one column per $\mathbf G_k$, $Q_i/101$ and $\sigma_e^2$. Five seeds per CSI level train 200000 steps in 200-slot terminal episodes; final checkpoints are evaluated. Historical four-candidate searches used two seeds and 50000 steps under power back-off, selecting SAC rate $3\times10^{-4}$/batch 256 and PPO rate $10^{-4}$/clip 0.1/four epochs/batch 256. SAC adapts entropy; PPO uses none. SE uses realized SINR. FIFO residence uses completed communication packets; the five-CSI runs leave $<0.026\%$ unfinished, and packet conservation holds.

\begin{table*}[t]
\caption{Confirmation performance at $\sigma_e^2=0.2$. Exceedance is mean$\pm$SD; other metrics are means. SD uses five streams for rules and five training-seed means for learned policies.}
\label{tab:main}
\centering\footnotesize\setlength{\tabcolsep}{5pt}
\begin{tabular}{lcccccc}\toprule
Metric & PPO & RLD-SAC & DPP & $\pi_0$ & $h$-myopic & \textbf{Anchored}\\\midrule
Task reward & 7.076 & 7.514 & 9.277 & 9.795 & 9.762 & \textbf{9.793}\\
Exceedance (\%) & $19.88\pm9.16$ & $11.43\pm1.49$ & $27.86\pm0.66$ & $3.62\pm0.17$ & $1.51\pm0.12$ & $\bm{2.63\pm0.06}$\\
SE (bit/s/Hz) & 6.012 & 5.517 & 6.535 & 6.330 & 6.269 & \textbf{6.263}\\
Sensing success (\%) & 83.90 & 86.74 & 98.15 & 98.15 & 97.78 & \textbf{98.15}\\
FIFO residence (ms) & 0.571 & 0.553 & 0.625 & 0.533 & 0.523 & \textbf{0.527}\\\bottomrule
\end{tabular}
\end{table*}

\begin{figure}[t]
\centering\includegraphics[width=\columnwidth]{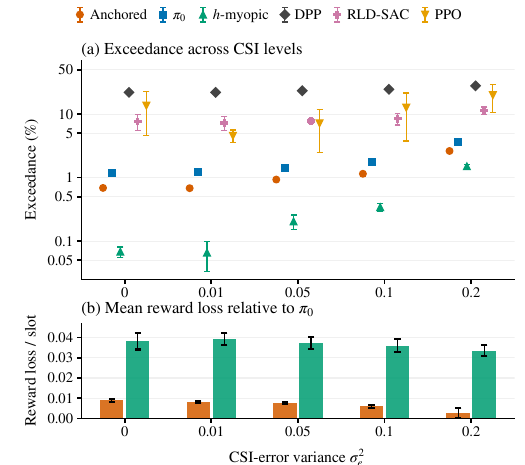}
\caption{(a) Exceedance (log scale); SDs follow Table~\ref{tab:main}. (b) Reward loss relative to $\pi_0$; SDs use five paired-stream differences. CSI levels use equally spaced positions.}
\label{fig:sweep}
\end{figure}

\subsection{Reward--Queue Trade-off}
\label{sub:tradeoff}
Table~\ref{tab:main} and Fig.~\ref{fig:sweep} place Anchored between $\pi_0$ and unclipped $h$-myopic control. Anchored lowers exceedance on all 25 paired streams relative to $\pi_0$; the paired reward losses in Fig.~\ref{fig:sweep}(b) average 0.0028--0.0089 per slot. Clipping preserves sensing success while accepting a smaller reduction in exceedance than the unbounded correction. The growth check (mean communication backlog $>100$, or last-2000-slot mean $>2$ times the first-2000-slot mean) passes on all 25 anchored confirmation streams. DPP's lowest validation exceedance across six weights is 15.4\%; all 50 learned checkpoints are outperformed by $\pi_0$ in reward and exceedance.

We retain $V=10$ as a common utility--queue compromise: relative to $V=5$, Anchored gains 0.087\,bit/s/Hz in SE for a 0.118-percentage-point exceedance increase (0.814\% to 0.932\%); at $V=20$, exceedance reaches 2.612\% (Fig.~\ref{fig:vsens}). The sensitivity check fixes $\chi=99$ and $\rho=0.5$.

\begingroup
\setlength{\intextsep}{6pt}
\begin{figure}[!htb]
\centering\includegraphics[width=\columnwidth,trim=0 11 0 0,clip]{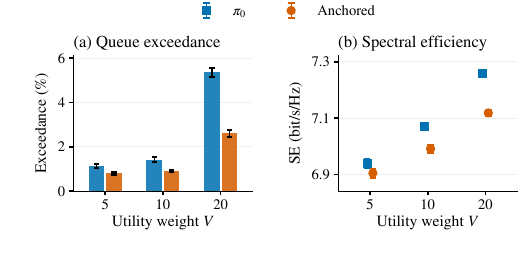}
\caption{Sensitivity to $V$ at $\sigma_e^2=0.05$: (a) exceedance and (b) SE. Both rules use five common streams; error bars show SDs. Anchored keeps $\chi=99$ and $\rho=0.5$ fixed.}
\label{fig:vsens}
\end{figure}
\endgroup

\subsection{Traffic and System-Scale Variation}
\label{sub:traffic}
At $\sigma_e^2=0.2$, scaling MMPP rates by $L=0.8,1,1.2$ gives exceedance reductions of 56.8/26.9/9.2\% with unchanged rule weights; absolute reductions grow from 0.150 to 1.932 percentage points. Unequal user-rate multipliers $(1+d,1+d,1-d,1-d)$, $d=0,0.25,0.5$, preserve total mean traffic and give reductions of 26.6/10.0/3.6\%. Thus relative gains decrease with load and imbalance, while Anchored lowers exceedance on every paired stream (15 per experiment). At $d=0.5$, the heavy-group mean/p99 backlog falls from 69.9/180 to 68.5/169 packets.

Table~\ref{tab:scale} fixes $N=8$, total power and total mean arrivals (500/3 packets and nine tasks per slot), scaling user/target rates by $4/M$ and $3/K$. Rules share nested pools of 2187/8192 ternary-logit candidates. Each scale retrains three seeds per learning method for 200000 steps, using full-power selected settings. Five common streams yield $5\times2\times3\times5=150$ policy--stream pairs. Anchored with 2187 candidates has higher reward and lower exceedance in every pair; the minimum reward gain is 1.492 per slot. These pairs share streams within each configuration. At 23 variables, expanding to 8192 candidates raises anchored reward from 8.971 to 9.078.

\begin{table}[t]
\caption{Scale-expansion exceedance (\%), $\sigma_e^2=0.2$, 2187 candidates. Rules average five streams; learning averages three five-stream seed means.}
\label{tab:scale}
\centering\footnotesize\setlength{\tabcolsep}{3pt}
\begin{tabular}{ccrrrr}\toprule
$(M,K)$ & $M+K$ & $\pi_0$ & RLD-SAC & PPO & \textbf{Anchored}\\\midrule
$(6,3)$ & 9 & 0.212 & 4.269 & 26.344 & \textbf{0.036}\\
$(8,3)$ & 11 & 0.016 & 10.980 & 42.612 & \textbf{0.000}\\
$(4,5)$ & 9 & 9.470 & 12.925 & 36.599 & \textbf{7.058}\\
$(8,7)$ & 15 & 0.522 & 21.835 & 48.161 & \textbf{0.072}\\
$(12,11)$ & 23 & 0.920 & 39.461 & 50.091 & \textbf{0.004}\\\bottomrule
\end{tabular}
\end{table}

\subsection{Retuning and Score Accuracy}
\label{sub:retuning}
A full-power search at $\sigma_e^2=0.2$ uses four configurations per learning method, two seeds and 200000 steps per run. Three validation streams select SAC rate $10^{-3}$/batch 512 and PPO rate $10^{-4}$/clip 0.05, at the rate/batch upper and clip lower endpoints. Five fresh seeds and five new common streams retain the reward/exceedance ordering: 7.520/8.41\% (SAC), 7.378/26.18\% (PPO), 9.800/3.74\% ($\pi_0$), 9.798/2.73\% (Anchored).

Per CSI level, 95 trajectory states, 2000 common error draws and a causal MMPP filter estimate a mean conditional gap of 0.017--0.029 for Anchored versus 0.0006--0.014 for $\pi_0$. The estimated $\rho+\osc(e)$ averages 0.64--1.03. Equal draw splits for selection and evaluation change mean gaps by $<10^{-4}$.

\subsection{Verified CPU Implementation}
\label{sub:timing}
Seven-variable fused C preserves candidate order with reference fallback. On one Xeon Gold 6148 core (2.40\,GHz), 200 warm-ups and 2000 calls per CSI level give median/p99 0.239--0.245/0.274--0.311\,ms, including input processing and action output. All 10000 timed actions match (maximum 0.388\,ms), as do all 25 complete trajectories (250000 slots; five fallbacks). Same-round actor p99 is 0.222--0.248\,ms. Compilation, CSI acquisition and control transmission are excluded.

\section{Conclusion}
The utility--backlog objective yields a task reward whose conditional value follows from integer-service moments. Anchored look-ahead refines its plug-in maximizer, reducing five-CSI exceedance by 27.4--43.9\% with mean reward loss below 0.01. Fixed-budget tests outperform freshly retrained learning baselines through 23 variables; the seven-variable implementation achieves p99 below 0.312\,ms.

\begingroup
\setlength{\abovedisplayskip}{5pt plus 1pt minus 1pt}
\setlength{\belowdisplayskip}{5pt plus 1pt minus 1pt}
\setlength{\abovedisplayshortskip}{3pt plus 1pt minus 1pt}
\setlength{\belowdisplayshortskip}{3pt plus 1pt minus 1pt}
\appendices
\section{Integer-Service Probabilities}
\label{app:service}
\subsection{Gaussian Moments and Departure Errors}
\label{app:gaussian}
For $\mathbf H=\mathbf H_{mj}$, centering the threshold form gives
\begin{equation}
X_{mj}-\bar x_{mj}=2\operatorname{Re}(\mathbf e_m^H\mathbf H\hat{\mathbf h}_m)
+\mathbf e_m^H\mathbf H\mathbf e_m-\sigma_e^2\operatorname{tr}\mathbf H.
\label{eq:centered}
\end{equation}
The linear and centered quadratic terms have variances $2\sigma_e^2\|\mathbf H\hat{\mathbf h}_m\|^2$ and $\sigma_e^4\operatorname{tr}(\mathbf H^2)$. Their covariance vanishes by the zero third centered Gaussian moments, proving~\eqref{eq:threshold_moments}. With $\epsilon_{ij}=p_{ij}-\indicator\{j\le\hat J_i\}$, integer tail sums also give
\begin{align}
\mathbb E[D_i-\hat J_i]&=\sum_{j=1}^{Q_i}\epsilon_{ij},\label{eq:deviation}\\
\mathbb E[(D_i-\hat J_i)^2]&=\sum_{j=1}^{Q_i}|2j-2\hat J_i-1|\,|\epsilon_{ij}|.\label{eq:deviation2}
\end{align}
Indeed, splitting at $d=\hat J_i$ yields $\sum_{j\le d}(2d-2j+1)(1-p_{ij})+\sum_{j>d}(2j-2d-1)p_{ij}$. Cantelli bounds $|\epsilon_{mj}|\le\psi_{mj}$, with $\psi_{mj}=s_{mj}^2/(s_{mj}^2+\bar x_{mj}^2)$ when $\bar x_{mj}>0$, $j\le\hat J_m$, or $\bar x_{mj}<0$, $j>\hat J_m$; otherwise $\psi_{mj}=1$. For sensing, $\psi_{ij}=0$.

\subsection{Spectral Series and Truncation}
\label{app:spectral}
On $\mathbf x\perp\mathbf w_{c,m}$, $\mathbf x^H\mathbf H_{mj}\mathbf x=-\tau_j\mathbf x^H\mathbf B_m\mathbf x\le0$, so $\mathbf H_{mj}$ has at most one positive eigenvalue. If it has none, $p_{mj}=0$ since $\tau_jn>0$. Otherwise, for $\sigma_e^2>0$, let $\lambda_+$ be its positive eigenvalue and $\bm\nu=\mathbf U^H\hat{\mathbf h}_m$ the estimate in its eigenbasis. For negative eigenvalues $\lambda_\ell$, define $q_+=|\nu_+|^2/\sigma_e^2$, $q_\ell=|\nu_\ell|^2/\sigma_e^2$, $\beta_\ell=-\lambda_\ell/\lambda_+$, $\theta_\ell=\beta_\ell/(1+\beta_\ell)$ and $\beta_0=\tau_jn/(\sigma_e^2\lambda_+)$. Failure is $T_+<Y$, where
\begin{equation}
T_+=|z_++\nu_+/\sigma_e|^2,\quad
Y=\beta_0+\sum_\ell\beta_\ell|z_\ell+\nu_\ell/\sigma_e|^2,
\label{eq:spectral_event}
\end{equation}
with independent standard complex Gaussian $z$'s. The noncentral exponential $T_+$ is a Poisson$(q_+)$ mixture of Gamma$(k+1,1)$ laws, giving
\begin{equation}
1-p_{mj}=\sum_{k\ge0}e^{-q_+}\frac{q_+^k}{k!}f_k,\qquad
f_k=1-\sum_{l=0}^kr_l,
\label{eq:series}
\end{equation}
with $r_l=\mathbb E[e^{-Y}Y^l/l!]$ and $u_\ell(z)=1+\beta_\ell(1-z)$, the generating function is
\begin{equation}
\begin{aligned}
R(z)&=\mathbb E[e^{-(1-z)Y}]=\sum_{l\ge0}r_lz^l,\\
&=e^{-\beta_0(1-z)}\prod_\ell\frac{e^{-q_\ell\beta_\ell(1-z)/u_\ell(z)}}{u_\ell(z)}.
\end{aligned}
\label{eq:generating}
\end{equation}
Writing $\log R(z)=\log r_0+\sum_{s\ge1}c_sz^s$ gives
\begin{equation}
\begin{aligned}
r_0&=e^{-\beta_0-\sum_\ell\theta_\ell q_\ell}\prod_\ell(1+\beta_\ell)^{-1},\\
c_s&=\beta_0\indicator\{s=1\}+\sum_\ell\theta_\ell^s[1/s+(1-\theta_\ell)q_\ell].
\end{aligned}
\label{eq:series_coefficients}
\end{equation}
Equating coefficients in $R'=R(\log R)'$ yields
\begin{equation}
r_l=\frac1l\sum_{s=1}^lsc_sr_{l-s},\qquad l\ge1.
\label{eq:recurrence}
\end{equation}
Here $c_s\ge0$, $r_l\ge0$ and $\sum_l r_l=R(1)=1$, so $0\le f_{k+1}=f_k-r_{k+1}\le f_k\le1$. Truncation after $L$ obeys
\begin{equation}
0\le\sum_{k>L}e^{-q_+}\frac{q_+^k}{k!}f_k
\le f_L\Pr\{\operatorname{Poisson}(q_+)>L\}.
\label{eq:series_remainder}
\end{equation}
At zero error variance, service thresholds are deterministic as stated in Theorem~\ref{thm:departures}.

\section{Score Error, Exceedance and Drift}
\label{app:score}
\subsection{Conditional Reward and Score Error}
\label{app:conditional}
Each communication departure depends only on its user's current error; sensing departures are known. Conditional independence and the queue update give $\mathbb E\bar q=\varphi$ and $\mathbb E\bar q^2=\varphi^2+\sigma_A^2+v_D$. Substitution into~\eqref{eq:reward} proves Proposition~\ref{prop:interface}; $\sigma_A^2$ includes conditional arrival covariances. Subtracting~\eqref{eq:plugin} gives~\eqref{eq:error}, where
\begin{equation}
d_\varphi=\sum_i\varpi_i(\lambda_i-\bar\lambda_i)-\sum_i\varpi_i\mathbb E[D_i-\hat J_i].
\label{eq:mean_error}
\end{equation}
The departure identities imply $|\mathbb E[D_i-\hat J_i]|\le\sum_j\psi_{ij}$ and $\operatorname{Var}D_i\le\sum_j|2j-2\hat J_i-1|\psi_{ij}$. The spectral series evaluates the moments exactly. The action-independent $\sigma_A^2$ cancels from $\osc(e)$.

\subsection{Exceedance and Drift Identities}
\label{app:risk}
Conditioning on $D_i=d$ and using $\bar F_i(k)-\bar F_i(k+d)=\sum_{j=1}^d b_i(k+j)$ proves~\eqref{eq:risk}. Set $b_i(k)=0$ for $k<0$; the sum is empty at $Q_i=0$. Independent user errors and traffic processes give the product for any-queue risk.

For $L(\mathbf Q)=\tfrac12\sum_iQ_i^2$, expanding $\mathbf Q'=\mathbf Q-\mathbf D+\mathbf A_t$ gives
\begin{equation}
\begin{aligned}
\Delta_L(\mathbf a)&=C_t(\mathbf a)+\sum_iQ_i(\lambda_i-\mathbb E[D_i]),\\
&\le C_t(\mathbf a)+\sum_iQ_i\big(\lambda_i-\hat J_i+\sum_{j\le\hat J_i}\psi_{ij}\big),
\end{aligned}
\label{eq:drift}
\end{equation}
where $C_t(\mathbf a)=\tfrac12\mathbb E[\|\mathbf A_t-\mathbf D\|^2\mid\mathcal F_t,\mathbf a]$ retains its action dependence. The inequality drops excess service and bounds predicted-departure shortfall~\cite{ref13_neely,ref_ouyang}.

\endgroup

\end{document}